\documentclass[twocolumn]{autart}    

\usepackage{graphicx}          
 \usepackage{mathtools}
 \usepackage{amssymb}
 \usepackage{balance}
 \usepackage{float}
\usepackage[scr]{rsfso}
 
 \newtheorem{assumption}{Assumption}
  \newtheorem{definition}{Definition}
  \usepackage{soul}	
 \newcommand{\Lim}[1]{\raisebox{0.5ex}{\scalebox{0.8}{$\displaystyle \lim_{#1}\;$}}}
\makeatletter
\def\@opargbegintheorem#1#2#3{\trivlist
   \item[]{\bfseries #1\ #2\ (#3)} \itshape}
\makeatother

\begin{document}

\begin{frontmatter}

\title{Efficiency Analysis of the Simplified Refined Instrumental Variable Method 
for Continuous-time Systems
\thanksref{footnoteinfo}} 

\thanks[footnoteinfo]{This paper was not presented at any IFAC meeting.}

\author[First]{Siqi Pan}\ead{siqi.pan@uon.edu.au},      
\author[First]{James S. Welsh}\ead{james.welsh@newcastle.edu.au},  
\author[Second]{Rodrigo A. Gonz\' alez}\ead{grodrigo@kth.se}  \;and	
\author[Second]{Cristian R. Rojas}\ead{crro@kth.se}  

\address[First]{School of Electrical Engineering and Computing, University 
of Newcastle, Callaghan, NSW 2308, Australia}  
\address[Second]{Division of Decision and
Control Systems, School of Electrical Engineering and Computer Science,
KTH Royal Institute of Technology, Stockholm 10044, Sweden}             


\begin{abstract}                          
In this paper, we derive the asymptotic Cram\'er-Rao lower bound for the continuous-time output 
error model structure and provide an analysis of the statistical efficiency of the Simplified 
Refined Instrumental Variable method for Continuous-time systems (SRIVC) 
based on sampled data.
It is shown that the asymptotic Cram\'er-Rao lower bound is independent of the 
intersample behaviour of the noise-free system output and hence only depends on the 
intersample behaviour of the system input.
We have also shown that, at the converging point of the SRIVC algorithm, 
the estimates do not depend on the intersample behaviour of the measured output. 
It is then proven that the SRIVC estimator is asymptotically efficient for the 
output error model structure under mild conditions. 
Monte Carlo simulations are performed to verify the asymptotic Cram\'er-Rao lower 
bound and the asymptotic covariance of the SRIVC estimates.
\end{abstract}

\end{frontmatter}

\section{Introduction}

Dynamical systems in the physical world are most often continuous-time (CT) in nature, thus it is  
more intuitive to obtain mathematical descriptions of these systems in terms of CT models  
than discrete-time (DT) models. 
There are two main approaches for identifying CT systems, namely the direct and indirect 
approaches~\cite{Rao2006}. The indirect approach first identifies a DT model and then 
transforms it to a CT model. The direct approach, on the other hand, identifies a 
CT model directly from sampled data.
Two well known algorithms in direct CT identification are the Refined Instrumental 
Variable method for Continuous-time system (RIVC) and its simplest embodiment, the Simplified 
RIVC (SRIVC) method \cite{Young1980}.

Despite the success that the RIVC and SRIVC algorithms have gained in practical 
applications in recent years \cite{Young2011,Young2006a}, 
there has been a lack of theoretical support available for them. 
One of the challenges that arises when estimating a CT model is that only 
sampled input-output data are available as measurements. Hence, directly estimating a CT model 
requires the measured data to be interpolated in some manner. It is noted that an incorrect 
interpolation of the sampled signals may lead to unsatisfactory estimates of the system parameters. 
The importance of intersample behaviour in CT system identification has been 
highlighted in the consistency analysis of the SRIVC estimator~\cite{Pan2019}, which shows 
that in order for the SRIVC estimator to be generically consistent, the intersample 
behaviour of the model input must match that of the system input.
However, most of the comments made 
with respect to the statistical efficiency of the estimators are based on either empirical 
observations~\cite{Young1980,Young2011,Garnier2008,Young2008,Young2002} 
or theorems developed for DT estimation methods~\cite{Young2015,Garnier2015,Chen2017}, 
which have overlooked the importance of the intersample behaviour of the signals in 
CT system identification as part of the theoretical analysis.



A desirable property for a consistent estimator to possess is statistical efficiency. In addition, 
a consistent estimator is asymptotically unbiased if there is a uniform upper bound on the variance. 
An unbiased estimator is then said to be efficient if its covariance 
matrix achieves the Cram\'er-Rao lower bound (CRLB) \cite{Goodwin1977,Ljung1999}.
With respect to the efficiency analysis of direct CT estimators such as SRIVC or 
RIVC, the existing literature \cite{Garnier2008,Garnier2007,Chen2013,Garnier2015} have claimed 
that the RIVC and SRIVC estimators are asymptotically efficient with additive coloured noise and white 
noise on the output respectively. 
Even though the CRLB expressions for the RIVC and SRIVC estimators exist in the literature, 
these expressions have been obtained directly through the results for DT instrumental 
variable algorithms developed in~\cite{Soderstrom1981,Stoica1983,Soderstrom1989}. 
Hence, a formal derivation 
where the intersample behaviour of the measured signals has been carefully addressed 
does not exist. In fact, the CRLB expression 
stated in~\cite[p. 105]{Garnier2008} is dependent on the filtered version of the sampled noise-free 
system output, which means that the entries in the existing CRLB depend on the interpolation of the 
output signal prior to the continuous-time filtering operations, e.g., assuming a zero-order hold (ZOH) on the 
output will yield a different CRLB expression than assuming a first-order hold (FOH). Hence, the 
existing results in the literature are not satisfactory.


The current paper is focused on the efficiency analysis of the SRIVC estimator.
The objectives of this paper are to derive an expression of the asymptotic CRLB for the continuous-time 
output error model structure and to examine the statistical efficiency of the SRIVC estimator. 
It is shown that the asymptotic CRLB is independent of the 
intersample behaviour of the noise-free system output and hence only depends on the 
intersample behaviour of the system input.
In the following sections, we will employ the notion of 
a theoretical SRIVC estimator, in which the measured output is considered as a CT signal 
purely for derivation purposes.
Since only sampled data are available as measurements, the practical implementation of the SRIVC 
algorithm requires an assumption on the intersample behaviour of the system output, such as 
a FOH, in order to perform the filtering operations. It has been shown in~\cite{Pan2019} that the effect of 
this intersample behaviour assumption on the output does not impact on the consistency of the 
SRIVC estimator. It is proven in the current paper that, at the converging point 
of the iterative algorithm, the standard SRIVC estimator is equivalent to the theoretical SRIVC estimator 
for large sample size. Using this fact, it is then shown that the asymptotic covariance of the SRIVC 
estimates coincides with the asymptotic CRLB when the intersample behaviour of the input signal 
in the regressor and instrument vectors matches that of the system input. The SRIVC estimator 
is therefore proven to be asymptotically efficient.

This paper is organised as follows. Section~\ref{sec:pre} provides the system and model definitions as well 
as a brief outline of the SRIVC estimator and the definition of the Cram\'er-Rao lower bound. This is followed 
by Section~\ref{sec:results} where the theoretical results of the paper, which includes theorems on the expressions 
for the asymptotic Cram\'er-Rao lower bound and the asymptotic covariance matrix of the SRIVC estimates, 
are presented. Section~\ref{sec:sim} provides simulation results that support the theoretical analysis, 
and the paper is concluded in Section 5.

\section{Preliminaries}	\label{sec:pre}

In this section, we define the structures of the continuous-time single-input single-output system and model, 
as well as provide a brief description of the SRIVC estimator and the definition of the Cram\'er-Rao lower bound.

\subsection{System and Model Definitions}

Consider a linear time-invariant CT system parameterised in the 
output error (OE) model structure
\begin{equation*}	\label{eq:true_sys}
 \mathcal{S}\colon \begin{cases}
               \mathring{x}(t) &= \dfrac{B^*(p)}{A^*(p)}\mathring{u}(t)\\
		y(t) &= \mathring{x}(t) + e(t),
            \end{cases}
\end{equation*}
where $p$ is the differential operator, i.e. $py(t)=\frac{d}{dt}y(t)$, $\mathring{x}(t)$ the 
noise-free system output and $\mathring{u}(t)$ the CT system input.
The system numerator and denominator polynomials are assumed coprime with orders given 
by $m^*$ and $n^*$ respectively, i.e.
\begin{equation*}	
\begin{split}
	B^*(p) &= b_0^*p^{m^*} + b_1^*p^{m^*-1} + \cdots + b_{m^*}^*,	\\ 
	A^*(p) &= a_1^*p^{n^*} + a_2^*p^{n^*-1} + \dots + a_{n^*}^*p + 1,
\end{split}
\end{equation*}
with the system parameter vector given by
\begin{equation*}
	\theta^* := \left[\begin{array}{cccccc}
	a_1^* & \dots & a_{n^*}^* & b_0^* & \dots & b_{m^*}^*
	\end{array}\right]^\top.
\end{equation*}
Only sampled input-output signals, denoted by $u(t_k)$ and $y(t_k)$, are available 
as measurements. 
Note that $u(t_k)=\mathring{u}(t)$ at the sampling instants, and the different notation indicates
that $u(t_k)$ may be interpolated with a different intersample behaviour from that of $\mathring{u}(t)$.
Due to the difficulty of dealing with the time-derivatives of stochastic noise, 
which does not have a finite variance, and the DT nature of the sampled signals, 
we only consider DT i.i.d. Gaussian noise given by $e\sim\mathcal{N}(0,\lambda)$. 
The output observation equation is then 
\begin{equation*}
	y(t_k) = \left\{\frac{B^*(p)}{A^*(p)} \mathring{u}(t)\right\}_{t=t_k} + e(t_k).
\end{equation*}

The model of the CT system is also parameterised as a proper 
transfer function, i.e.
\begin{equation}	\label{eq:OE_model}
 \mathcal{M}\colon \begin{cases}
               x(t_k) &= \dfrac{B(p)}{A(p)} u(t_k)\\
               y(t_k) &= x(t_k) + \varepsilon(t_k),
            \end{cases}
\end{equation}
with coprime numerator and denominator polynomials defined as
\begin{equation*}	\label{eq:model_parameters}
\begin{split}
	B(p) &= b_0p^{m} + b_1p^{m-1} + \cdots + b_m,	\\
	A(p) &= a_1p^{n} + a_2p^{n-1} + \dots + a_np + 1,
\end{split}
\end{equation*}
and the model parameter vector is given by
\begin{equation*}
	\theta := \left[\begin{array}{cccccc}
	a_1 & \dots & a_n & b_0 & \dots & b_m
	\end{array}\right]^\top.
\end{equation*}

When a mixed notation of CT operators and DT data is encountered, such as in~\eqref{eq:OE_model}, 
it is implied that the input~$u(t_k)$ is interpolated in some manner~\cite{Garnier2008}, e.g., using either 
a ZOH or a FOH, and the resultant output is sampled at $t_k$.



\subsection{The SRIVC Estimator}

The iterative SRIVC estimator minimises the sum of squares of the generalised equation 
error (GEE) \cite{Young1981}, 
$\varepsilon(t_k)$, which is given by
\begin{align} \label{eq:gee}
	\varepsilon(t_k) &= y(t_k) - x(t_k) 	\notag \\
				& = y(t_k) - \frac{B(p)}{A(p)}u(t_k)	\notag \\
				&= \frac{1}{A(p)}(A(p)y(t_k) - B(p)u(t_k))	\notag \\
				&= A(p)y_f(t_k) - B(p)u_f(t_k), \notag
\end{align}
where
\begin{equation} \label{eq:prefilter}
	y_f(t_k) = \frac{1}{A(p)} y(t_k), \; \text{and} \; u_f(t_k) = \frac{1}{A(p)} u(t_k).
\end{equation}

Due to the iterative nature of the SRIVC method, the $(j+1)$-th iteration of the SRIVC 
estimate~\cite{Young1980,Garnier2008} based on the parameter values estimated with sample size $N$ 
in the $j$-th iteration is given by
\begin{align}	\label{eq:srivc}
	\theta_{j+1}^N &= 
	\left[\frac{1}{N} \sum_{k=1}^N \hat{\varphi}_f(t_k,\theta_j^N) \varphi_f^\top(t_k,\theta_j^N) \right]^{-1} \notag\\
	 &\hspace{2cm}
	 \left[\frac{1}{N} \sum_{k=1}^N \hat{\varphi}_f(t_k,\theta_j^N) y_f(t_k,\theta_j^N) \right],
\end{align}
where the filtered regressor vector is
\begin{align}		\label{eq:srivc_reg}
	\varphi_f(t_k,\theta_j^N)
	&= \frac{1}{A_j(p)} \Big[\begin{array}{ccc}
	-p^n y(t_k) & \dots & -p y(t_k) 
	\end{array}\notag\\
	&\hspace{2.5cm}\begin{array}{ccc}
	p^mu(t_k) & \dots & u(t_k)
	\end{array}\Big]^\top, 	
\end{align}
and the filtered instrument vector is
\begin{align}		\label{eq:srivc_ins} 
	\hat{\varphi}_f(t_k,\theta_j^N) 
	&= \frac{1}{A_j(p)} \Big[\begin{array}{ccc}
	-\frac{B_j(p)}{A_j(p)} p^nu(t_k) & \dots & -\frac{B_j(p)}{A_j(p)}pu(t_k)
	\end{array}\notag\\
	&\hspace{2.6cm}\begin{array}{ccc}
	p^mu(t_k) & \dots & u(t_k)
	\end{array}\Big]^\top. 	
\end{align}
Note that to emphasise the dependency on the iteration~$j$ and the sample size $N$, the notation of the 
filtered output, $y_f(t_k)$, in~\eqref{eq:prefilter} is replaced by $y_f(t_k,\theta_j^N)$ in~\eqref{eq:srivc}.

The algorithm is stopped either when a maximum number of iterations is reached or when the 
relative error between the previous and current estimate is smaller than a constant, i.e.
\begin{equation}	\label{eq:srivc_stop}
	\frac{\|\theta^N_{j+1}-\theta^N_j\|}{\|\theta^N_{j+1}\|} < \epsilon.
\end{equation}

\vspace{-0.2cm}
Next, we define a ``theoretical'' SRIVC estimator where the system output is treated as a CT signal. 
This estimator is used purely for derivation purposes in the theoretical 
results section. 

\begin{definition}[Theoretical SRIVC estimator]	\label{def:theoretical_srivc}
Define a theoretical SRIVC estimator given by~\eqref{eq:srivc}, where the filtered regressor 
vector in~\eqref{eq:srivc_reg} is replaced by
\begin{align} \label{eq:lem_reg}
	\mathring{\varphi}_f(t_k,\theta_j^N) &= \Big[\begin{array}{cc}
	-\left\{\frac{p^n}{A_j(p)}\frac{B^*(p)}{A^*(p)}\mathring{u}(t) \right\}_{t=t_k} 
	+ \frac{p^n}{A_j(p)}e(t_k)& \dots 
	\end{array}\notag \\
	&\;\;\;\;\;\;\;\begin{array} {c}
	-\left\{\frac{p}{A_j(p)}\frac{B^*(p)}{A^*(p)}\mathring{u}(t) \right\}_{t=t_k}
	+ \frac{p}{A_j(p)}e(t_k) 
	\end{array}\notag\\
	&\;\;\;\;\;\;\;\begin{array}{ccc}
	\left\{\frac{p^m}{A_j(p)}\mathring{u}(t)\right\}_{t=t_k} & \dots 
	&\left\{\frac{1}{A_j(p)}\mathring{u}(t)\right\}_{t=t_k}
	\end{array}\Big]^\top,  
\end{align}

\vspace{-0.5cm}
and the filtered output in~\eqref{eq:prefilter} is replaced by
\begin{equation}	\label{eq:lem_yf}
	\mathring{y}_f(t_k,\theta_j^N) = \left\{\frac{1}{A_j(p)}\frac{B^*(p)}{A^*(p)}
	\mathring{u}(t)\right\}_{t=t_k} + \frac{1}{A_j(p)}e(t_k).
\end{equation}
Note that~\eqref{eq:lem_reg} and~\eqref{eq:lem_yf} implicitly assume that the 
measured output is a continuous-time signal. 
Thus, this estimator is not implemented in practice.
\end{definition}


\subsection{Cram\'er-Rao Lower Bound}

When assessing the performance of an estimator, the requirement of achieving a uniformly minimum mean 
square error is too stringent and unrealistic \cite{Goodwin1977}. Therefore, we concentrate on the class of 
asymptotically unbiased estimators (in the number of samples $N$) of the unknown parameter $\theta^*$, 
and on the \emph{asymptotic covariance matrix} of those estimators given by
\begin{equation}	\label{eq:def_ascov}
	\text{AsCov}(\theta^N) := \lim_{N\rightarrow\infty} N\mathbb{E}\{(\theta^N-\theta^*)(\theta^N-\theta^*)^\top\},
\end{equation}
whose trace coincides with the normalised asymptotic mean square error of the estimate 
$\theta^N$ due to its asymptotic unbiasedness. It can be shown 
(see	~\cite[Theorem 2.6, p.440]{Lehmann1998}) that, under mild conditions,
the asymptotic covariance of an estimator~\eqref{eq:def_ascov} is lower bounded, in a positive 
semi-definite sense, by the inverse of the Fisher information matrix per sample given by

\vspace{-0.8cm}
\scriptsize
\begin{equation}	\label{eq:cr_eq}
	P_{CR} = \left[\lim_{N\rightarrow\infty}
	\hspace{-0.05cm}\frac{1}{N}
	\hspace{-0.05cm}\left.\mathbb{E}\left\{\frac{\partial \log p(y^N;\theta)}{\partial \theta}
	\left(\frac{\partial \log p(y^N;\theta)}{\partial \theta}\right)^\top
	\right\}\right|_{\theta=\theta^*}
	\right]^{-1}\hspace{-0.3cm},
\end{equation}
\normalsize
for all values of $\theta^*$, except for a set of Lebesgue measure zero. 
We call~\eqref{eq:cr_eq} the \emph{asymptotic Cram\'er-Rao lower bound} on 
$\text{AsCov}(\theta^N)$. 
In~\eqref{eq:cr_eq}, the expression inside the inverse is known as the Fisher information matrix 
per sample, and $p(y^N,\theta)$ is the probability density function (PDF) of the full data $y^N$, 
which is parameterised by the vector $\theta$.
Note that the probability density function can be viewed as the likelihood function 
when expressed as a function of the parameter vector~$\theta$.

\subsection{Small-o Notation}

Next, a definition of the small-o notation for stochastic variables~\cite{vanderVaart2000}
is given in order to provide short expressions for terms that converge in probability to zero 
to facilitate the proof of the asymptotic distribution in 
Section~\ref{sec:asy_dis}.

\begin{definition}[Small-o notation] \label{def:small-o}
Let $X_N$ and $R_N$ be two sequences of random variables, then
\begin{equation*}
	X_N = o_p(R_N) \hspace{0.2cm} \text{means} \hspace{0.2cm}
	X_N = Y_NR_N \hspace{0.2cm} \text{and} \hspace{0.2cm} Y_N\overset{p}{\to} 0,
\end{equation*}
i.e. the sequence $X_N$ converges in probability to zero at the rate $R_N$.
\end{definition}


\section{Theoretical Results}	\label{sec:results}

In this section, we derive an expression of the asymptotic CRLB for the OE model structure and 
provide a theorem that describes the asymptotic distribution of the SRIVC estimates. The 
covariance expression of the estimates is then compared with the asymptotic CRLB to examine the 
statistical efficiency of the SRIVC estimator.

We begin by stating the assumptions required by the theorems developed in this section:

\begin{assumption}	\label{A1}
The system, $\frac{B^*(p)}{A^*(p)}$, is proper 
($n^* \geq m^*$) and asymptotically stable with $A^*(p)$ and $B^*(p)$ being coprime.
\end{assumption}
\begin{assumption}	\label{A2}
The input sequence, $u(t_k)$, and disturbance, $e(t_s)$, are stationary and mutually independent for all 
$k$ and $s$.
\end{assumption}
\begin{assumption}	\label{A3}
The input sequence, $u(t_k)$, is persistently exciting of order no less than $2n+1$.
\end{assumption}
\begin{assumption}	\label{A4}
All the zeros of $A_j(p)$ have strictly negative real parts, $n \geq m$, with $A_j(p)$ and $B_j(p)$ 
being coprime.
\end{assumption}
\begin{assumption}	\label{A5}
The model order matches the system order, i.e. $n=n^*$ and $m=m^*$.
\end{assumption}
\begin{assumption}	\label{A6}
The intersample behaviour of the input, $\mathring{u}(t)$, applied to the system is known 
exactly.
\end{assumption}

%

\subsection{Asymptotic Cram\'er-Rao Lower Bound}

In this subsection, we develop an explicit expression of the asymptotic CRLB for the continuous-time 
OE model structure. It is shown that the derived expression is independent of the 
intersample behaviour of the noise-free system output and hence only depends on the 
intersample behaviour of the system input.


\begin{thm} [Asymptotic Cram\'er-Rao lower bound] \label{theorem:CRLB}
Consider the prediction error 
\begin{equation}	\label{eq:prediction_error}
	\varepsilon(t_k,\theta) = y(t_k) - \left\{\frac{B(p)}{A(p)} \mathring{u}(t)\right\}_{t=t_k}
\end{equation}
for an unknown parameter vector $\theta$ formed using the model coefficients of $A(p)$ and $B(p)$.
Assume the output observations come from an output error model structure, i.e.
\begin{equation*}
	y(t_k) = \left\{\frac{B^*(p)}{A^*(p)} \mathring{u}(t)\right\}_{t=t_k} + e(t_k),
\end{equation*}
where $e(t_k)$ is i.i.d. Gaussian with variance $\lambda$, and $B^*(p)$ and $A^*(p)$ are the 
system polynomials. Then, under Assumptions~\ref{A1}, \ref{A2}, and~\ref{A5}, 
the asymptotic Cram\'er-Rao lower bound is given by
\begin{equation}	\label{eq:crlb_theorem}
	P_{CR} = \lambda \mathbb{E}\left\{ \psi(t_k,\theta^*) \psi^\top(t_k,\theta^*) \right\}^{-1},
\end{equation}
where
\begin{align}	\label{eq:crlb_psi}
	\psi(t_k,\theta^*) &= \Big[\begin{array}{ccc}
	-\frac{p^{n^*}B^*(p)}{{A^*}^2(p)}\mathring{u}(t) & \dots & 
	-\frac{pB^*(p)}{{A^*}^2(p)}\mathring{u}(t) 
	\end{array}\notag\\
	&\hspace{2cm}\left.\begin{array}{ccc}
	\frac{p^{m^*}}{A^*(p)}\mathring{u}(t) &
	\dots &
	\frac{1}{A^*(p)}\mathring{u}(t)
	\end{array}\Big]^\top\right|_{t=t_k.}
\end{align}
\end{thm}

\begin{pf*} {\textit{Proof of Theorem~\ref{theorem:CRLB}}}
According to \eqref{eq:prediction_error}, the joint PDF of the measured output 
based on $N$ samples, denoted by $y^N$, 
and hence the likelihood of $y^N$ when viewed as a function of 
unknown parameters $\theta$, is~\cite[Lemma 5.1]{Ljung1999}
\begin{equation*}	\label{eq:crlb_likelihood}
	p(\theta;y^N) = \prod_{k=1}^N p_e(\varepsilon(t_k,\theta)).
\end{equation*}

The log-likelihood of the measurement is thus given by
\begin{equation*}
\begin{split}
	\mathcal{L}(\theta) &= \log p(\theta;y^N)	\\
			&= \sum_{k=1}^N \log p_e(\varepsilon(t_k,\theta))	\\
			&= C - \frac{1}{2\lambda}\sum_{k=1}^N \varepsilon^2(t_k,\theta) \\
			&= C - \frac{1}{2\lambda}\sum_{k=1}^N
	\left[y(t_k) - \left\{\frac{B(p)}{A(p)} \mathring{u}(t)\right\}_{t=t_k}\right]^2,
\end{split}
\end{equation*}
where $C$ is a constant.

According to~\cite[Lemma 3.2]{Monaco1988}, the procedures of linearisation and discretisation commute. 
We can therefore differentiate $\mathcal{L}(\theta)$ with respect to the CT parameters and 
then discretise the transfer functions according to the intersample behaviour of the input signal and the sampling 
period.
Now, differentiating the log-likelihood function with respect to the denominator coefficients of the model 
and then evaluating at the system parameters gives
\begin{equation*}
\begin{split}
	\left.\frac{\partial\mathcal{L(\theta)}}{\partial a_i} \right|_{\theta=\theta^*}&=
	-\frac{1}{2\lambda} \cdot 2 \sum_{k=1}^N 
	\left[y(t_k) - \left\{\frac{B^*(p)}{A^*(p)} \mathring{u}(t)\right\}_{t=t_k}\right]\\
	&\hspace{2cm}\cdot
	\left[\left\{\frac{p^{n^*+1-i}B^*(p)}{(A^*(p))^2} \mathring{u}(t)\right\}_{t=t_k}\right]	\\
	&= -\frac{1}{\lambda} \sum_{k=1}^N e(t_k)
	\left\{\frac{p^{n^*+1-i}B^*(p)}{(A^*(p))^2} \mathring{u}(t)\right\}_{t=t_k},
\end{split}
\end{equation*}
where $i=1,\dots,n^*$.

Similarly, differentiating the log-likelihood with respect to the numerator coefficients gives
\begin{equation*}
\begin{split}
	\left.\frac{\partial\mathcal{L(\theta)}}{\partial b_i} \right|_{\theta=\theta^*}&=
	-\frac{1}{2\lambda} \cdot 2 \sum_{k=1}^N 
	\left[y(t_k) - \left\{\frac{B^*(p)}{A^*(p)} \mathring{u}(t)\right\}_{t=t_k}\right]\\
	&\hspace{2.85cm}\cdot
	\left[-\left\{\frac{p^{m^*-i}}{A^*(p)} \mathring{u}(t)\right\}_{t=t_k}\right]	\\
	&= \frac{1}{\lambda} \sum_{k=1}^N e(t_k)
	\left\{\frac{p^{m^*-i}}{A^*(p)} \mathring{u}(t)\right\}_{t=t_k},
\end{split}
\end{equation*}
where $i=0,\dots,m^*$.

Hence,
\begin{equation*}
	\left.\frac{\partial\mathcal{L(\theta)}}{\partial \theta} \right|_{\theta=\theta^*} = 
	\frac{1}{\lambda} \sum_{k=1}^N e(t_k) \psi(t_k,\theta^*),
\end{equation*}
where $\psi(t_k,\theta^*)$ is given in~\eqref{eq:crlb_psi}.
Note that the signals in~\eqref{eq:crlb_psi} are treated as CT signals and are filtered by CT transfer 
functions prior to sampling. Hence, it does not depend on the intersample behaviour 
of the noise-free system output.

The Fisher Information matrix is then given by
\begin{equation*}
\begin{split}
	I_F &= 
	\mathbb{E}\left\{\left(\left.\frac{\partial\mathcal{L(\theta)}}{\partial \theta} \right|_{\theta=\theta^*}\right)
	\left(\left.\frac{\partial\mathcal{L(\theta)}}{\partial \theta} \right|_{\theta=\theta^*}\right)^\top\right\}\\
	&=\frac{1}{\lambda^2}\sum_{k=1}^N\sum_{s=1}^N \mathbb{E}\left\{
	 e(t_k) \psi(t_k,\theta^*) e(t_s) \psi^\top(t_s,\theta^*) \right\}.
\end{split}
\end{equation*}
Since $u(t_k)$ and $e(t_k)$ are independent by Assumption~2, and $e(t_k)$ is i.i.d. Gaussian 
noise with variance $\lambda$,
\begin{align}	\label{eq:fisher1}
	I_F&=\frac{1}{\lambda^2}\sum_{k=1}^N\sum_{s=1}^N \mathbb{E}\left\{
	 e(t_k)e(t_s)\right\} \mathbb{E}\left\{\psi(t_k,\theta^*) \psi^\top(t_s,\theta^*) \right\} \notag\\
	&= \frac{1}{\lambda^2}\sum_{k=1}^N \lambda
	 \mathbb{E}\left\{\psi(t_k,\theta^*) \psi^\top(t_k,\theta^*) \right\}.
\end{align}
Now, $\psi(t_k,\theta^*)$ is composed of stationary random processes, hence its ensemble average is equal 
to its time average as sample size approaches infinity~\cite[Appendix B.1]{Soderstrom1989}, i.e.
\begin{equation*}
	\mathbb{E}\left\{ \psi(t_k,\theta^*) \psi^\top(t_k,\theta^*) \right\} = 
	\lim_{N\rightarrow \infty} \frac{1}{N} \sum_{l=1}^N\psi(t_l,\theta^*) \psi^\top(t_l,\theta^*)
\end{equation*}
for every $k = 1,...,N$. 
Therefore, the asymptotic Cram\'er-Rao lower bound, given by the inverse of the Fisher information 
matrix per sample, can be expressed as
\begin{align}	\label{eq:P_CR}
	P_{CR} &= \left[\lim_{N\rightarrow\infty}\frac{1}{N}I_F\right]^{-1} \notag\\
	&= \left[\lim_{N\rightarrow\infty}\frac{1}{N}\lambda^{-1} N \mathbb{E}\left\{ \psi(t_k,\theta^*) \psi^\top(t_k,\theta^*) \right\}\right]^{-1} \notag\\
	&=\lambda \mathbb{E}\left\{ \psi(t_k,\theta^*) \psi^\top(t_k,\theta^*) \right\}^{-1} \notag
\end{align}
where $\psi(t_k,\theta^*)$ is given in \eqref{eq:crlb_psi}.
\hfill$\qed$
\end{pf*}

\subsection{Asymptotic Distribution of the SRIVC Estimates} \label{sec:asy_dis}

Next, we derive the asymptotic distribution of the SRIVC estimates for the OE model structure 
under Assumptions 1 - 6.

\begin{thm} [Asymptotic distribution of the SRIVC estimates]	\label{theorem:cov_srivc}
Consider the SRIVC estimator given in \eqref{eq:srivc} under the output error model structure. 
Suppose Assumptions 1 - 6 hold, and assume that the estimator is consistent. 
Let $\bar{\theta}^N$ be the converging point of the SRIVC estimator for a fixed sampled 
size $N$, that is, $\bar{\theta}^N := \Lim{j\rightarrow \infty} \theta_j^N$, 
which corresponds to model polynomials $\bar{A}(p)$ and $\bar{B}(p)$.
Then, the SRIVC estimate is asymptotically Gaussian distributed, i.e.
\begin{equation}	\label{eq:cov_srivc_theorem1}
	\sqrt{N}(\bar{\theta}^N-\theta^*)  \xrightarrow[]{dist.} \mathcal{N}(0,P_{SRIVC}),
\end{equation}
where the asymptotic covariance matrix is
\begin{equation}	\label{eq:cov_srivc_theorem2}
	P_{SRIVC} = \lambda \mathbb{E}\left\{\tilde{\varphi}_f(t_k,\theta^*)
	\tilde{\varphi}_f^\top(t_k,\theta^*)\right\}^{-1}
\end{equation}
with $\tilde{\varphi}_f(t_k,\theta^*)$ given by
\begin{align}	\label{eq:srivc_phi}
	\tilde{\varphi}_f(t_k,\theta^*) &= \Big[\begin{array}{ccc}
	-\left(\frac{p^{n^*}B^*(p)}{{A^*}^2(p)}\right) u(t_k)
	& \dots & -\left(\frac{pB^*(p)}{{A^*}^2(p)}\right) u(t_k)
	\end{array}	\notag \\
	&\hspace{1.5cm}
	\begin{array}{ccc}
	\frac{p^{m^*}}{A^*(p)}u(t_k)  & \dots 
	& \frac{1}{A^*(p)}u(t_k)
	\end{array}\Big]^\top
\end{align} 
and $\lambda$ being the variance of the discrete-time i.i.d. Gaussian additive output noise.
\end{thm}




\vspace{-0.5cm}
\begin{pf*} {\textit{Proof of Theorem~\ref{theorem:cov_srivc}}}
From~\eqref{eq:srivc}, we know that
\begin{align}	\label{eq:thetaN-theta}
	\bar{\theta}^N -\theta^* &= \left[\frac{1}{N} \sum_{k=1}^N 
	\hat{\varphi}_f(t_k,\bar{\theta}^N) \varphi_f^\top(t_k,\bar{\theta}^N) \right]^{-1} \notag\\
	&\hspace{2cm}
	 \left[\frac{1}{N} \sum_{k=1}^N \hat{\varphi}_f(t_k,\bar{\theta}^N) y_f(t_k,\bar{\theta}^N) \right] - \theta^* \notag\\
	 &= \left[\frac{1}{N} \sum_{k=1}^N 
	 \hat{\varphi}_f(t_k,\bar{\theta}^N) \varphi_f^\top(t_k,\bar{\theta}^N) \right]^{-1} \notag\\
	 &
	 \left[\frac{1}{N} \sum_{k=1}^N 
	 \hat{\varphi}_f(t_k,\bar{\theta}^N) (y_f(t_k,\bar{\theta}^N)-\varphi_f^\top(t_k,\bar{\theta}^N)\theta^*) \right].
\end{align}

\vspace{-0.8cm}
It has been first stated in Remark~5 of~\cite{Pan2019} that the intersample behaviour 
of $y(t_k)$ assumed in order to perform the filtering operations in the regressor 
vector does not affect the SRIVC estimates at the converging point of the algorithm. 
Furthermore, it is proven in Lemma~\ref{lemma1} (see Appendix) that the standard SRIVC 
estimator~\eqref{eq:srivc} is equivalent to the theoretical estimator given in 
Definition~\ref{def:theoretical_srivc} at the converging point for a large sample size. 
Hence, according to Lemma~\ref{lemma1}, at $\bar{\theta}^N$, the filtered regressor and 
the filtered output in~\eqref{eq:thetaN-theta} can be replaced by~\eqref{eq:lem_reg} 
and~\eqref{eq:lem_yf} evaluated at $\bar{\theta}^N$ respectively.
Note that $u(t_k)$ must have the same intersample behaviour as the system input in order to obtain 
consistent estimates according to~~\cite[Theorem 1]{Pan2019}, 
and $e(t_k)$ is the DT noise added to the system output.
The second half of~\eqref{eq:thetaN-theta} can then be expressed as
\begin{align}
	&\mathring{y}_f(t_k,\bar{\theta}^N)-\mathring{\varphi}_f^\top(t_k,\bar{\theta}^N)\theta^* \notag\\
	&\hspace{1cm}
	=\left\{\frac{A^*(p)}{\bar{A}(p)}\frac{B^*(p)}{A^*(p)}\mathring{u}(t)\right\}_{t=t_k}
	+\frac{A^*(p)}{\bar{A}(p)}e(t_k) \notag\\
	&\hspace{5cm} - \left\{\frac{B^*(p)}{\bar{A}(p)}\mathring{u}(t)\right\}_{t=t_k}
	\notag\\
	&\hspace{1cm}
	=\frac{A^*(p)}{\bar{A}(p)}e(t_k).	\notag
\end{align}	
Therefore, \eqref{eq:thetaN-theta} simplifies to
\begin{align}
\bar{\theta}^N -\theta^* &= \left[\frac{1}{N} \sum_{k=1}^N 
	 \hat{\varphi}_f(t_k,\bar{\theta}^N) \mathring{\varphi}_f^\top(t_k,\bar{\theta}^N) \right]^{-1}\notag\\
	 &\hspace{1.2cm}
	 \left[\frac{1}{N} \sum_{k=1}^N \hat{\varphi}_f(t_k,\bar{\theta}^N) 
	\frac{A^*(p)}{\bar{A}(p)}e(t_k) \right]. \notag
\end{align}

Now, the first-order Taylor series expansion of $\sqrt{N}(\bar{\theta}^N-\theta^*)$ 
can be written as
\scriptsize
\begin{align}
	&\sqrt{N}(\bar{\theta}^N-\theta^*) \notag\\
	 &= \left[\frac{1}{N} \sum_{k=1}^N \hat{\varphi}_f(t_k,\theta^*) 
	\mathring{\varphi}_f^\top(t_k,\theta^*) \right]^{-1}
	 \left[\frac{1}{\sqrt{N}} \sum_{k=1}^N \hat{\varphi}_f(t_k,\theta^*)e(t_k)\right]	\notag \\
	 &\hspace{1cm}
	 +\left.\frac{\partial}{\partial \bar{\theta}^N} \left\{\left[\frac{1}{N} 
	 \sum_{k=1}^N \hat{\varphi}_f(t_k,\bar{\theta}^N) 
	\mathring{\varphi}_f^\top(t_k,\bar{\theta}^N) \right]^{-1}\right\} 
	\right|_{\bar{\theta}^N=\theta^*}	\notag\\
	&\hspace{3.5cm}
	\left[\frac{1}{\sqrt{N}} \sum_{k=1}^N \hat{\varphi}_f(t_k,\theta^*)
	e(t_k) \right] (\bar{\theta}^N-\theta^*) 	\notag\\
	&
	 + \left[\frac{1}{N} \sum_{k=1}^N \hat{\varphi}_f(t_k,\theta^*) 
	\mathring{\varphi}_f^\top(t_k,\theta^*) \right]^{-1}\hspace{-0.15cm}
	\Bigg(\frac{1}{\sqrt{N}}\sum_{k=1}^N
	 \left.\frac{\partial \hat{\varphi}_f(t_k,\bar{\theta}^N)}
	 {\partial\bar{\theta}^N}\right|_{\bar{\theta}^N=\theta^*}\hspace{-0.6cm}e(t_k) 	\notag\\
	 &
	 + \frac{1}{\sqrt{N}}\sum_{k=1}^N \hat{\varphi}_f(t_k,\theta^*)
	 \left.\frac{\partial}{\partial \bar{\theta}^N}
	 \left(\frac{1}{\bar{A}(p)}\right)\right|_{\bar{\theta}^N=\theta^*}
	 \hspace{-0.2cm}A^*(p)
	 e(t_k)\Bigg)(\bar{\theta}^N-\theta^*) \notag	\\
	 &+ o_p(\sqrt{N}\|\bar{\theta}^N - \theta^*\|).	 \label{eq:big_eq}
\end{align}
\normalsize

Let
\begin{equation*}
	R(\theta) = \left[\frac{1}{N} \sum_{k=1}^N \hat{\varphi}_f(t_k,\theta) 
	\mathring{\varphi}_f^\top(t_k,\theta) \right]^{-1},
\end{equation*}
then~\eqref{eq:big_eq} can be rearranged to be
\scriptsize
\begin{subequations} \label{eq:big_eq_total}
\begin{align}
	&\sqrt{N} \Bigg\{ I - \left.\frac{\partial R(\theta)}{\partial \theta} \right|_{\theta=\theta^*}
	\left[\frac{1}{N} \sum_{k=1}^N \hat{\varphi}_f(t_k,\theta^*)e(t_k)\right]	\label{eq:big_eq1}\\
	& -R(\theta^*) \frac{1}{N}\sum_{k=1}^N
	 \left.\frac{\partial \hat{\varphi}_f(t_k,\bar{\theta}^N)}
	 {\partial\bar{\theta}^N}\right|_{\bar{\theta}^N=\theta^*}\hspace{-0.5cm}e(t_k)	\label{eq:big_eq2}\\
	 &  -R(\theta^*)\frac{1}{N}\sum_{k=1}^N \hat{\varphi}_f(t_k,\theta^*)
	 \left.\frac{\partial}{\partial \bar{\theta}^N}
	 \left(\frac{1}{\bar{A}(p)}\right)\right|_{\bar{\theta}^N=\theta^*}
	 \hspace{-0.8cm}A^*(p) e(t_k) + o_p(1)\Bigg\}(\bar{\theta}^N-\theta^*)	\label{eq:big_eq3}\\
	 &= R(\theta^*) \left[\frac{1}{\sqrt{N}} \sum_{k=1}^N \hat{\varphi}_f(t_k,\theta^*)e(t_k)\right], \label{eq:big_eq4}
\end{align}
\end{subequations}
\normalsize
where $I$ is the identity matrix.

\vspace{-0.5cm}
Next, we will examine the behaviours of~\eqref{eq:big_eq1} -- \eqref{eq:big_eq4} for large sample size. 
Since the estimator is consistent, the 
matrix inverse $R(\theta^*)$ in ~\eqref{eq:big_eq2} -- \eqref{eq:big_eq4} 
is non-singular~\cite{Pan2019} for sufficiently large $N$.
According to Assumption~\ref{A2}, the input and disturbance are stationary, then by the ergodic lemma 
in~\cite[Lemma 3.1]{Soderstrom1975}, the matrix product term in~\eqref{eq:big_eq1} can be written as
\begin{equation*}
\begin{split}
	&\left.\frac{\partial R(\theta)}{\partial \theta} \right|_{\theta=\theta^*}
	\left[\frac{1}{N} \sum_{k=1}^N \hat{\varphi}_f(t_k,\theta^*)e(t_k)\right]	\\
	& \hspace{1.5cm}= \left.\frac{\partial R(\theta)}{\partial \theta} \right|_{\theta=\theta^*}
	\mathbb{E}\left\{\hat{\varphi}_f(t_k,\theta^*) e(t_k)\right\} + o_p(1)
\end{split}
\end{equation*}
for large sample size, where $o_p(1)$ is the small-o notation given in Definition~\ref{def:small-o}.
Since the instrument vector consists of filtered inputs that are independent of the disturbance $e(t_k)$ 
according to Assumption~\ref{A2}, by using the same method as in the proof 
of~\cite[Theorem 1]{Pan2019}, 
we can show that $ \mathbb{E}\left\{\hat{\varphi}_f(t_k,\theta^*) e(t_k)\right\}=0$. 
Hence,
\begin{equation*}
	\left.\frac{\partial R(\theta)}{\partial \theta} \right|_{\theta=\theta^*}
	\left[\frac{1}{N} \sum_{k=1}^N \hat{\varphi}_f(t_k,\theta^*)e(t_k)\right]
	= o_p(1).
\end{equation*}

In addition, 
by adopting the numerator layout, i.e. the Jacobian formulation, when performing vector differentiation, we obtain
\begin{equation*}	\label{eq:diff1}
\left.\frac{\partial \hat{\varphi}_f(t_k,\bar{\theta}^N)}{\partial \bar{\theta}^N}\right|_{\bar{\theta}^N=\theta^*}  
= M(p)u(t_k),
\end{equation*}
where
\begin{equation*}
\begin{split}
&M(p) = \left(\frac{1}{A^*(p)}\right)^2\cdot\\
&\begin{bmatrix}
        2\frac{B^*(p)}{A^*(p)}p^{2n^*}  & \dots  &  2\frac{B^*(p)}{A^*(p)}p^{n^*+1}  
        &  -p^{n^*+m^*}  &  \dots  & -p^{n^*}	 \\
	 \vdots & & \vdots & \vdots &  & \vdots \\
       2\frac{B^*(p)}{A^*(p)}p^{n^*+1}  & \dots  &  2\frac{B^*(p)}{A^*(p)}p^{2}  &  -p^{m^*+1}  
        &  \dots  & -p	 \\
        \\
        -p^{n^*+m^*}  & \dots  & -p^{m^*+1}  &   0	&    \dots &    0	 \\
        \vdots & & \vdots & \vdots &  & \vdots \\
        -p^{n^*}  & \dots  & -p  &   0	&    \dots &    0	 
\end{bmatrix}.
\end{split}
\end{equation*}
For a large sample size and by the same reasoning, \eqref{eq:big_eq2} 
can be written as
\begin{equation*}
\begin{split}
	&R(\theta^*)\frac{1}{N} \sum_{k=1}^N 
	\left.\frac{\partial \hat{\varphi}_f(t_k,\bar{\theta}^N)}{\partial \bar{\theta}^N}
	\right|_{\bar{\theta}^N=\theta^*}e(t_k) 	\\
	&\hspace{2.5cm}= R(\theta^*)\mathbb{E}\{M(p) u(t_k) e(t_k)\} + o_p(1)	\\
	&\hspace{2.5cm} = o_p(1).
\end{split}
\end{equation*}

Furthermore,
\begin{equation*}
	\left.\frac{\partial}{\partial \bar{\theta}^N}\left(\frac{1}{\bar{A}(p)}\right)\right|_{\bar{\theta}^N=\theta^*}	
	= \frac{1}{{A^*}^2(p)}[-p^{n^*} \; \dots \; -p \; 0 \; \dots \; 0].
\end{equation*}
Then, for a large sample size, the product term inside the bracket of \eqref{eq:big_eq3} can be expressed as
\begin{equation*}
\begin{split}
	 &R(\theta^*)\frac{1}{N}\sum_{k=1}^N \hat{\varphi}_f(t_k,\theta^*)
	 \left.\frac{\partial}{\partial \bar{\theta}^N}\left(\frac{1}{A(p)}\right)
	 \right|_{\bar{\theta}^N=\theta^*}\hspace{-0,5cm}A^*(p)e(t_k)\\
	 &= R(\theta^*)\mathbb{E}\Big\{\hat{\varphi}_f(t_k,\theta^*) \frac{1}{A^*(p)}[-p^n\;\dots\;-p\;0\;\dots\;0]
	 e(t_k) \Big\}+ o_p(1)\\
	 &=o_p(1).
\end{split}
\end{equation*}

Now, \eqref{eq:big_eq1} -- \eqref{eq:big_eq3} is simplified to
\begin{equation}	\label{eq:big_eq_left}
	\sqrt{N}\left(I + o_p(1)\right)(\bar{\theta}^N-\theta^*),
\end{equation}
and \eqref{eq:big_eq_total} then becomes
\begin{equation*}
\begin{split}
	\sqrt{N}(\bar{\theta}^N-\theta^*) 
	&= \left(I + o_p(1)\right)^{-1}\left[\frac{1}{N} \sum_{k=1}^N \hat{\varphi}_f(t_k,\theta^*) 
	\mathring{\varphi}_f^\top(t_k,\theta^*) \right]^{-1} \\
	&\hspace{3cm}
	\left[\frac{1}{\sqrt{N}} \sum_{k=1}^N \hat{\varphi}_f(t_k,\theta^*)e(t_k)\right].
\end{split}
\end{equation*}

According to \cite{Soderstrom1975}, as the sample size approaches infinity,
\begin{equation*}
	\frac{1}{N} \sum_{k=1}^N \hat{\varphi}_f(t_k,\theta^*) 
	\mathring{\varphi}_f^\top(t_k,\theta^*) \rightarrow
	\mathbb{E}\left\{ \hat{\varphi}_f(t_k,\theta^*) 
	\mathring{\varphi}_f^\top(t_k,\theta^*) \right\}
\end{equation*}
with probability $1$, Hence, by~\cite[Lemma A4.3]{Soderstrom1983},
\begin{equation*}
	\left[\frac{1}{N} \sum_{k=1}^N \hat{\varphi}_f(t_k,\theta^*) 
	\mathring{\varphi}_f^\top(t_k,\theta^*) \right]^{-1} \hspace{-0.5cm}\overset{p}{\to} \hspace{-0.1cm}
	\mathbb{E}\left\{ \hat{\varphi}_f(t_k,\theta^*) 
	\mathring{\varphi}_f^\top(t_k,\theta^*) \right\}^{-1}\hspace{-0.4cm}.
\end{equation*}
Let $\tilde{\varphi}_f(t_k,\theta^*)$ in~\eqref{eq:srivc_phi} be the noise-free version of 
$\mathring{\varphi}_f(t_k,\theta_j)$ given in~\eqref{eq:lem_reg} evaluated at $\theta^*$.
Then, for large $N$,
\begin{equation*}
\begin{split}
	\sqrt{N}(\bar{\theta}^N-\theta^*) &= \left(I + o_p(1)\right)^{-1}\mathbb{E}\{ \hat{\varphi}_f(t_k,\theta^*) 
	\mathring{\varphi}_f^\top(t_k,\theta^*) \}^{-1} \\
	&\hspace{2.5cm}
	\left[\frac{1}{\sqrt{N}} \sum_{k=1}^N \hat{\varphi}_f(t_k,\theta^*)e(t_k)\right]\\
	&= \left(I + o_p(1)\right)^{-1}\mathbb{E}\{ \hat{\varphi}_f(t_k,\theta^*) \tilde{\varphi}_f^\top(t_k,\theta^*) \}^{-1}\\
	&\hspace{2.5cm}
	\left[\frac{1}{\sqrt{N}} \sum_{k=1}^N \hat{\varphi}_f(t_k,\theta^*)e(t_k)\right].
\end{split}
\end{equation*}


Since both $\hat{\varphi}(t_k)$ and $e(t_k)$ are stationary and independent, 
by~\cite[Lemma A4.1]{Soderstrom1983},
\begin{equation*}
	\frac{1}{\sqrt{N}} \sum_{k=1}^N \hat{\varphi}_f(t_k,\theta^*)e(t_k) \xrightarrow[]{dist.} \mathcal{N}(0,P),
\end{equation*}
where
\begin{align} \label{eq:P_ex3}
	P &= \lim_{N\rightarrow\infty} \frac{1}{N}\sum_{k=1}^{N}\sum_{s=1}^N
	\mathbb{E}\left\{[\hat{\varphi}_f(t_k,\theta^*)e(t_k)]
	[\hat{\varphi}_f(t_s,\theta^*)e(t_s)]^\top\right\} \notag \\
	&=\lim_{N\rightarrow\infty} \frac{1}{N}\sum_{k=1}^{N}\sum_{s=1}^N
	\mathbb{E}\left\{e(t_k)e(t_s)\right\}\mathbb{E}\left\{\hat{\varphi}_f(t_k,\theta^*)
	\hat{\varphi}_f^\top(t_s,\theta^*)\right\} \notag \\
	&= \lim_{N\rightarrow\infty} \frac{1}{N}\sum_{k=1}^{N} \lambda
	\mathbb{E}\left\{\hat{\varphi}_f(t_k,\theta^*)\hat{\varphi}_f^\top(t_k,\theta^*)\right\} \notag \\
	&= \lim_{N\rightarrow\infty} \frac{1}{N}\lambda N
	\mathbb{E}\left\{\hat{\varphi}_f(t_k,\theta^*)\hat{\varphi}_f^\top(t_k,\theta^*)\right\} \notag \\
	&= \lambda \mathbb{E}\left\{\hat{\varphi}_f(t_k,\theta^*)\hat{\varphi}_f^\top(t_k,\theta^*)\right\}.
\end{align}


Now, by~\cite[Lemma A4.2]{Soderstrom1983} and its corollary, and that $(I + o_p(1))^{-1} = I + o_p(1)$ 
(which follows from the continuous mapping theorem \cite[Theorem 2.3]{vanderVaart2000}),
this implies that
\begin{equation*}	
	\sqrt{N}(\bar{\theta}^N-\theta^*)  \xrightarrow[]{dist.} \mathcal{N}(0,P_{SRIVC}),
\end{equation*}
where
\begin{align}	\label{eq:cov_IV}
	P_{SRIVC} &= \mathbb{E}\left\{\hat{\varphi}_f(t_k,\theta^*)\tilde{\varphi}_f^\top(t_k,\theta^*)\right\}^{-1}P\notag\\
	&\hspace{1.5cm}
	\mathbb{E}\left\{\tilde{\varphi}_f(t_k,\theta^*)\hat{\varphi}_f^\top(t_k,\theta^*)\right\}^{-1},
\end{align}
with $P$ given in \eqref{eq:P_ex3}. Substituting \eqref{eq:P_ex3} into \eqref{eq:cov_IV}, we can express 
the asymptotic covariance matrix as
\begin{align}	\label{eq:cov_srivc_white}
	&P_{SRIVC} = \lambda \mathbb{E}\left\{\hat{\varphi}_f(t_k,\theta^*)\tilde{\varphi}_f^\top(t_k,\theta^*)\right\}^{-1}\notag\\
	&\hspace{0.5cm}
	\mathbb{E}\left\{\hat{\varphi}_f(t_k,\theta^*)  \hat{\varphi}^\top_f(t_k,\theta^*)\right\}
	\mathbb{E}\left\{\tilde{\varphi}_f(t_k,\theta^*)\hat{\varphi}_f^\top(t_k,\theta^*)\right\}^{-1}\hspace{-0.4cm},
\end{align}
where $\tilde{\varphi}_f(t_k,\theta^*)$ is given in \eqref{eq:srivc_phi}.

Note that if the intersample behaviours of the input signal in the instrument vector, 
in the form of~\eqref{eq:srivc_ins}, are chosen to be exactly the same as that of the input signal 
in the regressor vector, 
i.e. $\hat{\varphi}_f(t_k,\theta^*) = \tilde{\varphi}_f(t_k,\theta^*)$, 
then~\eqref{eq:cov_srivc_white} simplifies to

\hfill $P_{SRIVC} = \lambda \mathbb{E}\left\{\tilde{\varphi}_f(t_k,\theta^*)
	\tilde{\varphi}_f^\top(t_k,\theta^*)\right\}^{-1}.$ \qed
\end{pf*}

\begin{rem}~\label{rem:eff}
From Theorems~\ref{theorem:CRLB} and~\ref{theorem:cov_srivc}, \eqref{eq:srivc_phi} is equal 
to~\eqref{eq:crlb_psi} when the intersample behaviour of the model input matches that of the 
system input.
Hence, the asymptotic covariance of the SRIVC estimates in~\eqref{eq:cov_srivc_theorem2} 
coincides with the asymptotic CRLB in~\eqref{eq:crlb_theorem}.
We therefore conclude that the SRIVC estimator is asymptotically efficient under the output error model 
structure.
\end{rem}

\begin{rem}
The result in Remark~\ref{rem:eff} implies that in order for the SRIVC estimator to be asymptotically efficient, 
the instrument vector must be chosen in the same form as the noise-free regressor of the theoretical SRIVC 
estimator in~\eqref{eq:lem_reg}. This means that in the digital implementation of the SRIVC estimator, the 
two filtering operations for the generation of the model output, $x(t_k)$, and the filtered derivatives of the 
model output, $x_f^{(i)}(t_k)$, must be combined into a single filtering process to ensure the asymptotic 
efficiency of the SRIVC estimator. The filtered instrument vector in~\eqref{eq:srivc_ins}, which is usually how 
it is presented in the existing literature (see e.g.~\cite{Garnier2008}), can be rewritten 
more explicitly in the form of~\eqref{eq:srivc_phi}.
\end{rem}

\begin{cor}	\label{cor1}
In general, the SRIVC estimator is not asymptotically efficient if the intersample behaviour 
of the input in the filtered instrument vector in~\eqref{eq:srivc_ins} 
does not match that of the system input under the output error model structure.
\end{cor}

\begin{pf*} {\textit{Proof of Corollary~\ref{cor1}}}
The proof follows from~\cite[Lemma A3.9]{Soderstrom1983}, which states that
\begin{equation*}
\begin{split}
	&\mathbb{E}\{\hat{\varphi}_f(t_k)\tilde{\varphi}_f^\top(t_k)\}^{-1}
	\mathbb{E}\{\hat{\varphi}_f(t_k)\hat{\varphi}_f^\top(t_k)\}
	\mathbb{E}\{\tilde{\varphi}_f(t_k)\hat{\varphi}_f^\top(t_k)\}^{-1} \\
	&\hspace{4cm}
	\succeq \mathbb{E}\{\tilde{\varphi}_f(t_k)\tilde{\varphi}_f^\top(t_k)\}^{-1}.
\end{split}
\end{equation*}
Note that a strict inequality is achieved when there exists no invertible matrix that relates $\hat{\varphi}_f(t_k)$
to $\tilde{\varphi}_f(t_k)$. $\hfill\qed$
\end{pf*}

\section{Simulation Results}	\label{sec:sim}

In this section, Monte Carlo (MC) simulations are performed with both first and second order systems 
to provide empirical evidence to the theoretical results presented in the previous section. 

\subsection{Simulation 1: First order system} 

The first order system is chosen to be
\begin{equation*}
	G^*(p) = \frac{b_0^*}{a_1^*p+1},
\end{equation*}
where the parameters are given by $\theta^*=[a_1^*\; b_0^*]^\top=[0.1\; 10]^\top$. 
The input/output signals are sampled at~$T = 0.01$ sec, and the input is chosen to be an 
i.i.d. Gaussian sequence with a unity variance, which is then interpolated with a ZOH. 
The additive noise on the output is also an i.i.d. Gaussian sequence with a unity variance and 
is uncorrelated with the input.
The sample size, $N$, is $2\times10^5$, and $5\times10^4$ MC runs are 
performed with the SRIVC estimator initialised at $\theta^*$. 
The maximum number of iterations of the SRIVC algorithm is set to 200, and the 
relative error bound in~\eqref{eq:srivc_stop} is set to $10^{-12}$.
The covariance 
of the asymptotic distribution of the SRIVC estimate given in~\eqref{eq:cov_srivc_theorem2} 
is then approximated using $5\times10^4$ sets of estimates, giving

\vspace{-1cm}
\small
\begin{equation}	\label{eq:1st_cov}
	P_{SRIVC} \approx \left[\begin{array}{cc}
	(8.0327\pm0.0508) \times 10^{-3} & 0.3996\pm0.0021	\\
	0.3996\pm0.0021 & 39.8223\pm0.2519
	\end{array}\right],
\end{equation}
\normalsize
where the mean and standard deviation of each covariance entry in~\eqref{eq:1st_cov} are determined using 
the method outlined in~\cite[Appendix B.9]{Soderstrom1989}.

The expectation of two signals filtered by CT transfer functions can be computed by first converting the 
CT transfer functions to their DT ZOH equivalents and then using the method outlined 
in~\cite[Section~5]{Soderstrom2003}. The CRLB in~\eqref{eq:crlb_psi}
can thus be computed analytically at $\theta^*$ to be
\begin{align}	\label{eq:1st_crlb}
	P_{CR} 
	= \left[\begin{array}{cc}
	8.0334 \times 10^{-3} & 0.4010	\\
	0.4010 & 40.0333
	\end{array}\right],
\end{align}
where~\eqref{eq:1st_crlb} is rounded to four decimal places. 
The analytical expression of the CRLB in~\eqref{eq:1st_crlb} matches well with the approximated 
covariance matrix in~\eqref{eq:1st_cov} being within the standard deviations.

Now, it has been stated in the existing 
literature~\cite{Garnier2008,Garnier2007,Chen2013,Garnier2015} that the SRIVC estimator 
is asymptotically efficient with the covariance matrix, and therefore also the CRLB, defined in the 
same form as~\eqref{eq:cov_srivc_theorem2} but with the filtered regressor given by 
(see e.g. \cite[p. 105]{Garnier2008})
\begin{align}	\label{eq:reg_lit}
	&\tilde{\varphi}_f(t_k) \notag\\
	&= \frac{1}{A^*(p)}\left[\begin{array}{cccccc}
	\mathring{x}^{(n^*)}(t_k) & \dots & \mathring{x}^{(1)}(t_k) & u^{(m^*)}(t_k) & \dots & u(t_k)
	\end{array}\right]^\top \notag \\
	&= \frac{1}{A^*(p)}\left[\begin{array}{cccccc}
	p^{n^*}\mathring{x}(t_k) & \dots & p\mathring{x}(t_k) & p^{m^*}u(t_k) & \dots & u(t_k)
	\end{array}\right]^\top,
\end{align}
where $\mathring{x}(t_k)$ is the sampled version of the noise-free system output.
It can be seen that~\eqref{eq:reg_lit} is different to the filtered regressor defined in~\eqref{eq:srivc_phi} in 
Theorem~\ref{theorem:cov_srivc}. Computing the covariance matrix of the SRIVC estimator or the 
CRLB expression using~\eqref{eq:reg_lit} will lead to the wrong results as the intersample behaviour of 
the noise-free system output $\mathring{x}(t_k)$ is not captured correctly since only sampled signals 
are available in practice. 
The key difference between the CRLB and covariance expressions derived in this paper and the expressions 
given in the existing literature 
is that the evaluation of the filtered regressor in~\eqref{eq:srivc_phi} implicitly assumes that the 
noise-free system output is a CT signal, whereas~\eqref{eq:reg_lit} assumes the noise-free 
output $\mathring{x}(t_k)$ is interpolated in some manner where a mixed notation of CT transfer 
function and sampled data is used~\cite[p. 96]{Garnier2008}.
For instance, assuming a ZOH for $\mathring{x}(t_k)$ during 
the filtering operations when computing the covariance matrix from the existing literature~\cite{Garnier2008} 
will result in
\begin{equation*}	
	P_{SRIVC}^{lit} = \left[\begin{array}{cc}
	7.2629 \times 10^{-3} & 0.3813	\\
	0.3813 & 40.0333
	\end{array}\right],
\end{equation*}
which does not match the covariance matrix approximated using the MC simulations 
in~\eqref{eq:1st_cov}.

\begin{figure} [h]
\begin{center}
\includegraphics[width = 8.5cm]{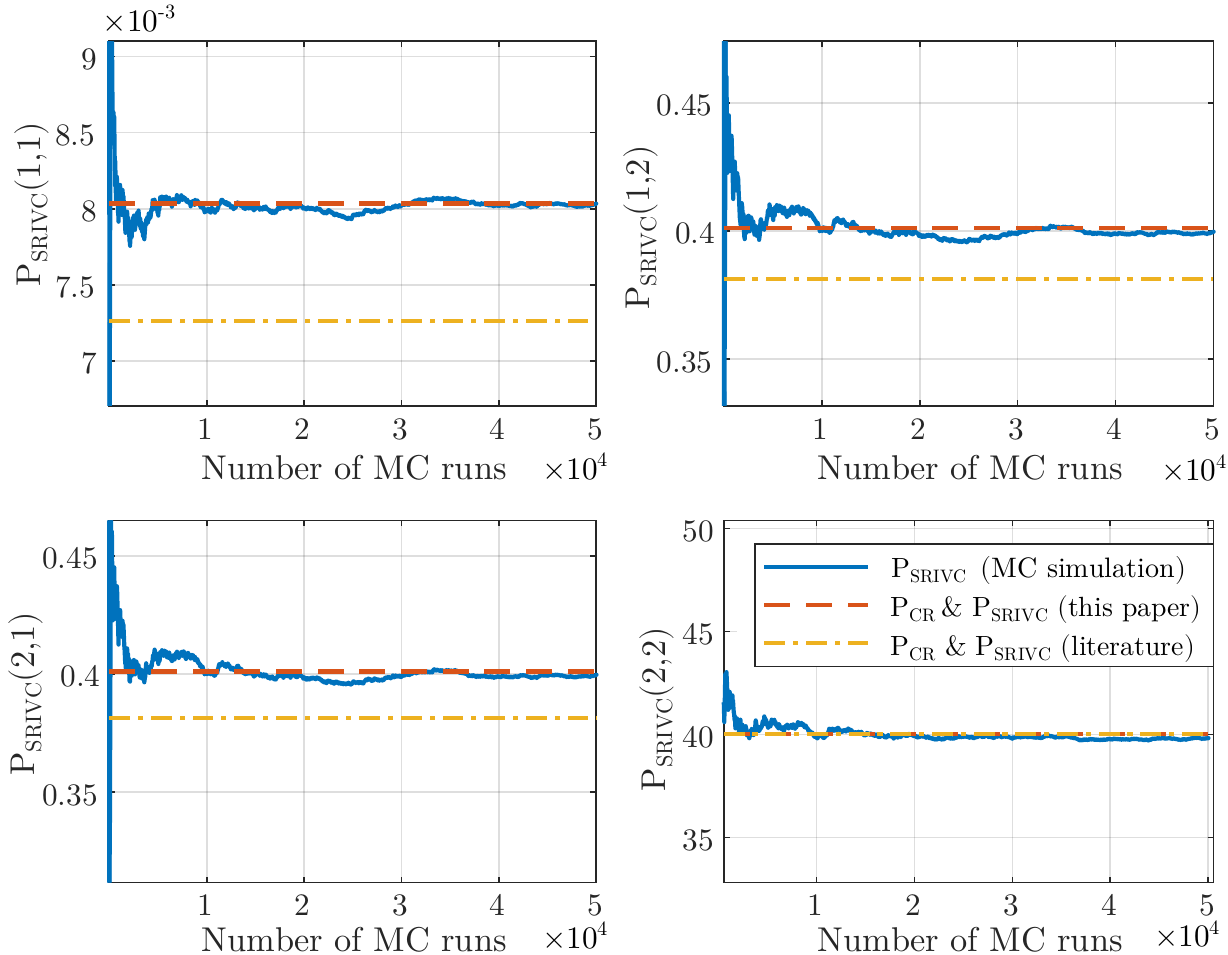}
\caption{Covariance of the first order transfer function estimates.}
\label{fig:1stOrder_cov_fit}
\end{center}
\end{figure}

The covariance matrix approximated through the MC simulations is plotted against the 
number of MC runs as shown in Figure~\ref{fig:1stOrder_cov_fit}.
The CRLB/covariance derived in this paper and the CRLB/covariance in the existing literature are 
also shown in Figure~\ref{fig:1stOrder_cov_fit}. It can 
be seen that the covariance matrix obtained in simulation converges to the CRLB derived in this paper, 
which also coincides with the derived covariance matrix of the SRIVC estimates, with an increasing number 
of MC runs. This provides empirical evidence that the SRIVC estimator is asymptotically efficient under 
the output error model structure. 
However, the CRLB or the covariance expression in the existing literature does not match 
the simulation results.

\subsection{Simulation 2: Second order system}

\begin{figure} [h]
\begin{center}
\includegraphics[width = 8.5cm]{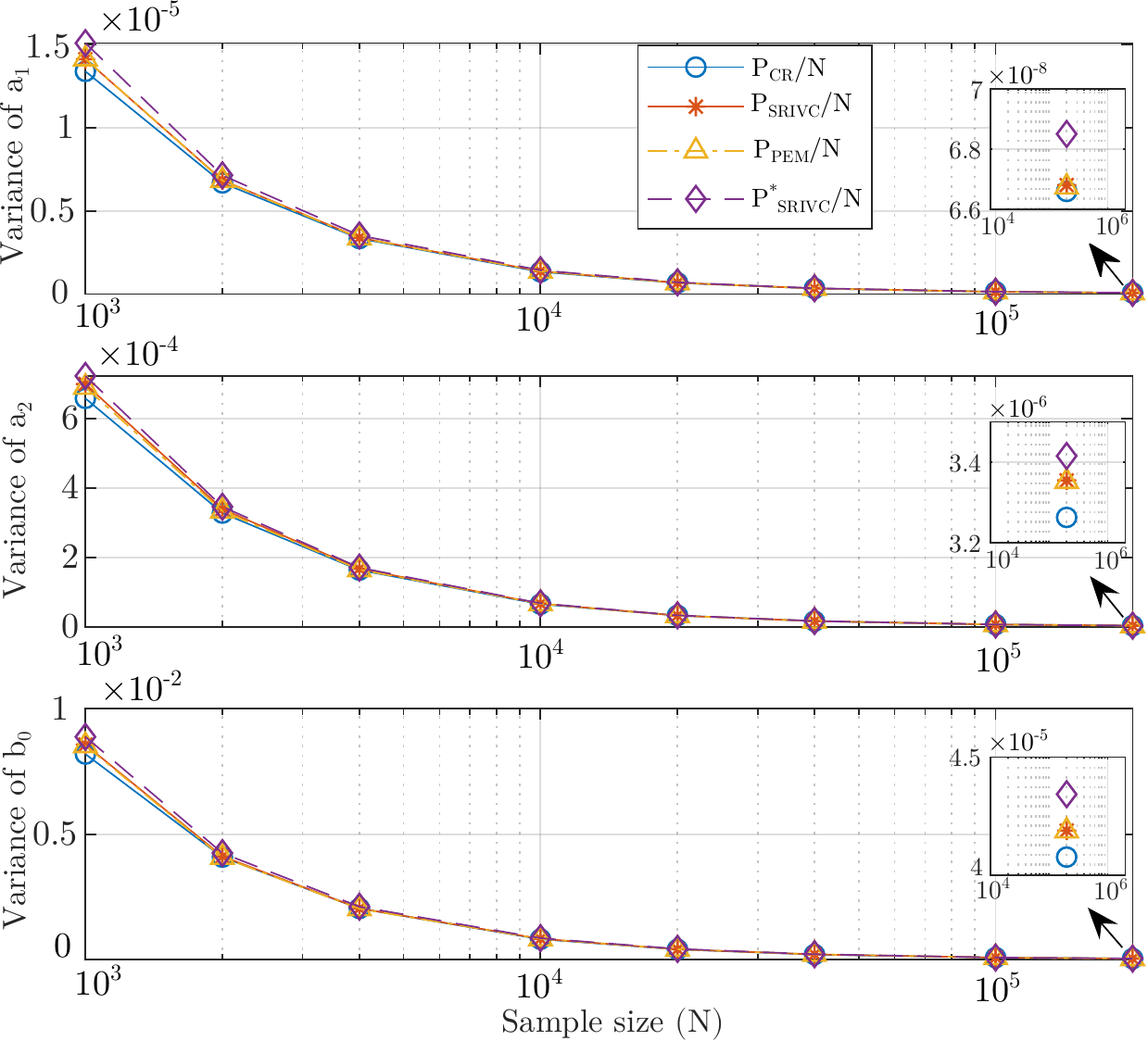}
\caption{Variance of the second order transfer function estimates.}
\label{fig:2ndOrder}
\end{center}
\end{figure}

The second order system is chosen to be
\begin{equation*}
	G^*(p) = \frac{b_0^*}{a_1^*p^2+a_2^*p+1},
\end{equation*}
where the parameters are given by 
$\theta^*=[a_1^*\; a_2^*\; b_0^*]^\top=[0.04\; 0.2\; 1]^\top$. The input and additive noise
settings as well as the relative error bound and maximum iterations of the SRIVC algorithm 
are the same as the settings for Simulation 1. The signals are sampled at~$T=0.1$ sec,
the sample size $N$ is varied from $10^3$ to $2\times10^5$ in eight steps, and 
$10^4$ MC simulations are performed for each $N$ with the SRIVC estimator initialised at $\theta^*$. 
The covariance matrix of the estimates are then approximated from the distributions generated by the 
MC simulations, and the three diagonal entries are plotted against $N$ in Figure~\ref{fig:2ndOrder}.
In addition to the CRLB calculated using~\eqref{eq:crlb_theorem}, the system parameters are also 
estimated with an asymptotically efficient indirect PEM method proposed in~\cite{Gonzalez2018} 
for comparison as shown in Figure~\ref{fig:2ndOrder}. 
This method enforces a fixed relative degree in the estimated CT transfer function, 
and its statistical efficiency has been shown in~\cite{Gonzalez2018}.
Furthermore, the fourth instance in Figure~\ref{fig:2ndOrder} corresponds to the situation where 
the SRIVC estimator uses a FOH input signal in the filtered instrument vector with the covariance 
matrix labelled as $P^*_{SRIVC}/N$.

It can be observed in Figure~\ref{fig:2ndOrder} that the variance of the SRIVC estimates converges 
quickly to the CRLB with an increasing sample size, and the small discrepancies are due to the finite 
sample approximation of the covariance matrix. The variance of the SRIVC estimates are 
indistinguishable from that of the asymptotically efficient indirect PEM estimates as seen 
in the upper right windows (zoom view of the last sample size)
 in Figure~\ref{fig:2ndOrder}. These provide empirical evidence to the 
asymptotic efficiency of the SRIVC estimator. On the other hand, when the intersample 
behaviour of the input signal in the filtered instrument vector does not match the system input in the 
SRIVC estimator, a higher variance of the estimates can be observed, which indicates that the 
estimator is not efficient in this case and it aligns with Corollary~\ref{cor1}.

\vspace{-0.1cm}

\vspace{-0.2cm}
\section{Conclusion}	\label{sec:conclusion}

\vspace{-0.2cm}
In this paper, we have derived the asymptotic Cram\'er-Rao lower bound for the continuous-time output error 
model structure and provided the asymptotic covariance expression of the SRIVC estimator by explicitly 
incorporating the intersample behaviour of the signals as part of the analysis. 
The asymptotic CRLB and the covariance expression 
derived in this paper are both different to the results reported in the existing literature.
It has been shown that the asymptotic CRLB is independent of the 
intersample behaviour of the noise-free system output and hence only depends on the 
intersample behaviour of the system input.
It has also been shown that the standard SRIVC estimator is equivalent to a theoretical SRIVC 
estimator at the converging point for large sample size, and this result has been employed 
to derive the asymptotic covariance of the SRIVC estimates. We conclude that the 
SRIVC estimator is asymptotically efficient under the output error model structure, 
i.e. the asymptotic covariance expression coincides with the 
asymptotic CRLB, when the intersample behaviour of the input signal in both the filtered regressor and 
instrument vectors matches that of the system input. 

\vspace{-0.2cm}
\section{Acknowledgements}		\label{sec:acknowledgments}

\vspace{-0.1cm}
This work was in part supported by the Australian government Research Training Program (RTP) scholarship, 
and in part by the Swedish Research Council under contract
number 2016-06079 (NewLEADS).

\section*{Appendix}

\begin{lem} [Equivalence between practical and theoretical SRIVC estimators]	\label{lemma1}
Consider the SRIVC iterations in \eqref{eq:srivc} for finite $N$ with $\varphi_f(t_k,\theta_j^N)$, 
$\hat{\varphi}_f(t_k,\theta_j^N)$ and $y_f(t_k,\theta_j^N)$ 
defined in~\eqref{eq:srivc_reg}, \eqref{eq:srivc_ins} and~\eqref{eq:prefilter} respectively. 
Let the converging point of the iterations be $\bar{\theta}^N:= \Lim{j\to \infty} \theta_j^N$, 
and assume the matrix 
\begin{equation}	\label{eq:lemma_normal}
\frac{1}{N}\sum_{k=1}^{N}\hat{\varphi}_f(t_k,\bar{\theta}^N)\varphi_f^\top(t_k,\bar{\theta}^N)
\end{equation}
is non-singular. 
Also consider the theoretical SRIVC estimator given in Defintion~\ref{def:theoretical_srivc} at 
the converging point~$\bar{\theta}^N$, and further assume that~\eqref{eq:lemma_normal} 
remains non-singular with the filtered regressor replaced by~\eqref{eq:lem_reg}. 
Then, there exists an integer $N_0$ such that 
the SRIVC estimator given in~\eqref{eq:srivc} is equivalent to the theoretical SRIVC estimator 
given in Definition~\ref{def:theoretical_srivc} evaluated at the converging point for $N>N_0$, 
that is, \eqref{eq:srivc} at the converging point can be expressed as
\begin{equation*}
\begin{split}
	\bar{\theta}^N &= 
	\left[\frac{1}{N} \sum_{k=1}^N \hat{\varphi}_f(t_k,\bar{\theta}^N) 
	\mathring{\varphi}_f^\top(t_k,\bar{\theta}^N) \right]^{-1} \\
	&\hspace{3cm}
	 \left[\frac{1}{N} \sum_{k=1}^N \hat{\varphi}_f(t_k,\bar{\theta}^N) 
	 \mathring{y}_f(t_k,\bar{\theta}^N) \right].
\end{split}
\end{equation*}
\end{lem}

\begin{pf*} {\textit{Proof of Lemma~\ref{lemma1}}}
The converging point $\bar{\theta}$ of the SRIVC estimator must satisfy~\eqref{eq:srivc}, i.e.
\begin{equation*}
\begin{split}
	\bar{\theta}^N &= \left[\frac{1}{N}\sum_{k=1}^{N}\hat{\varphi}_f(t_k,\bar{\theta}^N)
	\varphi_f^\top(t_k,\bar{\theta}^N)\right]^{-1} \\
	&\hspace{3cm}
	\left[\frac{1}{N} \sum_{k=1}^{N}\hat{\varphi}_f(t_k,\bar{\theta}^N) y_f(t_k,\bar{\theta}^N)\right]. 
\end{split}
\end{equation*}
Equivalently,
\begin{align} \label{equivalently}
	&\frac{1}{N} \sum_{k=1}^{N}\hat{\varphi}_f(t_k,\bar{\theta}^N) y_f(t_k,\bar{\theta}^N)\notag\\
	&\hspace{1.5cm}
	-\frac{1}{N}\sum_{k=1}^{N}\hat{\varphi}_f(t_k,\bar{\theta}^N)\varphi_f^\top(t_k,\bar{\theta}^N)\bar{\theta}^N 
	= 0.
\end{align}
Note that 
\begin{equation*}
\begin{split}
	&y_f(t_k,\bar{\theta}^N) - \varphi_f^\top(t_k,\bar{\theta}^N)\bar{\theta}^N 	\\
	&= \frac{1}{\bar{A}(p)}y(t_k) - \frac{-\bar{a}_1 p^n - 
	\cdots - \bar{a}_n p}{\bar{a}_1 p^n + \cdots + \bar{a}_n p+1}y(t_k) 
	+ \frac{\bar{B}(p)}{\bar{A}(p)}u(t_k), 
\end{split}
\end{equation*}
where the filtering on $y(t_k)$ depends on the hold reconstruction that is chosen. 
Therefore, \eqref{equivalently} reduces to
\begin{align}\label{srivcconverging}
\frac{1}{N}& \sum_{k=1}^{N}\hat{\varphi}_f(t_k,\bar{\theta}^N) \left(y(t_k) 
- \frac{\bar{B}(p)}{\bar{A}(p)}u(t_k)  \right) = 0,
\end{align}
which does not depend on the hold chosen for $y(t_k)$. 

Now, consider the theoretical SRIVC estimator evaluated at the converging point 
with the filtered regressor and the filtered output given by~\eqref{eq:lem_reg} 
and~\eqref{eq:lem_yf}, we have
\begin{align}
	&\mathring{y}_f(t_k,\bar{\theta}) - \mathring{\varphi}_f^\top(t_k,\bar{\theta})\bar{\theta}\notag\\
	&\hspace{1cm}
	= \frac{B^*(p)}{\bar{A}(p)A^*(p)}u(t_k) + \frac{1}{\bar{A}(p)}e(t_k) \notag\\
	&\hspace{1.5cm}
	+ \frac{(\bar{a}_1p^n +\cdots+ \bar{a}_np)B^*(p)}{\bar{A}(p)A^*(p)}u(t_k) \notag\\
	&\hspace{2.3cm}
	+ \frac{\bar{a}_1p^n +\cdots+ \bar{a}_np}{\bar{A}(p)}e(t_k)	
	-\frac{\bar{B}(p)}{\bar{A}(p)}u(t_k)	\notag\\
	&\hspace{1cm}
	= \frac{\bar{A}(p)B^*(p)}{\bar{A}(p)A^*(p)}u(t_k) + \frac{\bar{A}(p)}{\bar{A}(p)}
	e(t_k) - \frac{\bar{B}(p)}{\bar{A}(p)}u(t_k)\notag\\
	&\hspace{1cm}
	=y(t_k) - \frac{\bar{B}(p)}{\bar{A}(p)}u(t_k),	\notag
\end{align}
which gives the same expression as the standard SRIVC estimator in~\eqref{srivcconverging}.
This means that both estimators solve the same equation for the parameters at the 
converging point for any $N>0$.
Following the proof in Statement~2 of Theorem~1 in~\cite{Pan2019}, it can be shown that there is a 
unique converging point for the SRIVC estimator as the sample size approaches infinity.
Therefore, replacing the filtered regressor vector and filtered output in the SRIVC estimator 
by~\eqref{eq:lem_reg} and~\eqref{eq:lem_yf} will result in the same estimate $\bar{\theta}$ 
for large sample size.
\hfill$\qed$
\end{pf*}

\balance
\bibliographystyle{plain}        
\bibliography{library}

\begin{thebibliography}{10}

\bibitem{Chen2013}
F.~Chen, H.~Garnier, and M.~Gilson.
\newblock Refined instrumental variable identification of continuous-time {OE}
  and {BJ} models from irregularly sampled data.
\newblock In {\em Adaptation and Learning in Control and Signal Processing},
  volume~11, pages 80--85, 2013.

\bibitem{Chen2017}
F.~Chen, M.~Gilson, H.~Garnier, and T.~Liu.
\newblock Robust time-domain output error method for identifying
  continuous-time systems with time delay.
\newblock {\em Systems and Control Letters}, 102:81--92, 2017.

\bibitem{Garnier2015}
H.~Garnier.
\newblock Direct continuous-time approaches to system identification. overview
  and benefits for practical applications.
\newblock {\em European Journal of Control}, 24:50 -- 62, 2015.

\bibitem{Garnier2007}
H.~Garnier, M.~Gilson, P.~C. Young, and E.~Heselstein.
\newblock An optimal {IV} technique for identifying continuous-time transfer
  function model of multiple input systems.
\newblock {\em Control Engineering Practice}, 15:471 -- 486, 2007.

\bibitem{Garnier2008}
H.~Garnier and L.~Wang, editors.
\newblock {\em Identification of Continuous-time Models from Sampled data}.
\newblock Springer, 2008.

\bibitem{Gonzalez2018}
R.~A. Gonz\'{a}lez, C.~R. Rojas, and J.~S. Welsh.
\newblock An asymptotically optimal indirect approach to continuous-time system
  identification.
\newblock In {\em IEEE Conference on Decision and Control (CDC)}, pages 638 --
  643, 2018.

\bibitem{Goodwin1977}
G.~C. Goodwin and R.~L. Payne.
\newblock {\em Dynamic system identification: Experiment design and data
  analysis}.
\newblock Academic Press, 1977.

\bibitem{Lehmann1998}
E.~L. Lehmann and G.~Casella.
\newblock {\em Theory of point estimation}.
\newblock Springer, 2nd edition, 1998.

\bibitem{Ljung1999}
L.~Ljung.
\newblock {\em System identification: Theory for the user}.
\newblock Prentice Hall, 2nd edition, 1999.

\bibitem{Monaco1988}
S.~Monaco and D.~Normand-Cyrot.
\newblock Zero dynamics of sampled nonlinear systems.
\newblock {\em Systems \& Control Letters}, 11(3):229 -- 234, 1988.

\bibitem{Pan2019}
S.~Pan, R.~A. Gonz\'{a}lez, J.~S. Welsh, and C.~R. Rojas.
\newblock Consistency analysis of the simplified refined instrumental variable
  method for continuous-time systems.
\newblock {\em Automatica}, 113, March 2020.

\bibitem{Rao2006}
G.~P. Rao and H.~Unbehauen.
\newblock Identification of continuous-time systems.
\newblock {\em IEE Proceedings - Control Theory and Applications}, 153(2):185
  -- 220, 2006.

\bibitem{Soderstrom1975}
T.~S\"oderstr\"om.
\newblock Ergodicity results for sample covariances.
\newblock {\em Problems of Control and Information Theory}, 4(2):131--138,
  1975.

\bibitem{Soderstrom2003}
T.~S\"oderstr\"om.
\newblock Computational methods for evaluating covariance functions.
\newblock Technical report, Systems and Control Group, Uppsala University,
  2003.

\bibitem{Soderstrom1981}
T.~S\"oderstr\"om and P.~Stoica.
\newblock Comparison of some instrumental variable methods -- consistency and
  accuracy aspects.
\newblock {\em Automatica}, 17(1):101--115, 1981.

\bibitem{Soderstrom1983}
T.~S\"oderstr\"om and P.~Stoica.
\newblock {\em Instrumental Variable Methods for System Identification}.
\newblock Springer-Verlag, 1983.

\bibitem{Soderstrom1989}
T.~S\"oderstr\"om and P.~Stoica.
\newblock {\em System Identification}.
\newblock Prentice Hall, 1989.

\bibitem{Stoica1983}
P.~Stoica and T.~S\"oderstr\"om.
\newblock Optimal instrumental variable estimation and approximate
  implementations.
\newblock {\em IEEE Transactions on Automatic Control}, 28(7):757--772, 1983.

\bibitem{vanderVaart2000}
A.~W. van~der Vaart.
\newblock {\em Asymptotic Statistics}.
\newblock Cambridge University Press, 2000.

\bibitem{Young1981}
P.~C. Young.
\newblock Parameter estimation for continuous-time models -- a survey.
\newblock {\em Automatica}, 17(1):23--39, 1981.

\bibitem{Young2002}
P.~C. Young.
\newblock Optimal {IV} identification and estimation of continuous-time {TF}
  models.
\newblock In {\em 15th IFAC World Congress}, volume~35, pages 109--114, 2002.

\bibitem{Young2008}
P.~C. Young.
\newblock The refined instrumental variable method: Unified estimation of
  discrete and continuous-time transfer function models.
\newblock {\em Journal Europeen des Systemes Automatises}, 42:149--179, 2008.

\bibitem{Young2011}
P.~C. Young.
\newblock {\em Recursive estimation and time-series analysis: An introduction
  for the student and practitioner}.
\newblock Springer-Verlag, 2nd edition, 2011.

\bibitem{Young2015}
P.~C. Young.
\newblock Refined instrumental variable estimation: Maximum likelihood
  optimization of a unified box-jenkins model.
\newblock {\em Automatica}, 52:35--46, 2015.

\bibitem{Young2006a}
P.~C. Young and H.~Garnier.
\newblock Identification and estimation of continuous-time, data-based
  mechanistic models for environmental systems.
\newblock {\em Environmental Modelling \& Software}, 21(8):1055--1072, 2006.

\bibitem{Young1980}
P.~C. Young and A.~J. Jakeman.
\newblock Refined instrumental variable methods of recursive time-series
  analysis part {III}. {E}xtensions.
\newblock {\em International Journal of Control}, 31(4):741--764, 1980.

\end{thebibliography}

\end{document}